\documentclass[preprint,prl,superscriptaddress,showpacs]{revtex4}
\usepackage{xspace}
\usepackage{graphicx}
\usepackage{color}
\usepackage{amsmath, amsthm, amssymb}
\usepackage{bm}

\newcommand{\ucsbphysics}{Department of Physics, University of California,
                          Santa Barbara, California 93106, USA}
\newcommand{\ucsbchem}{Department of Chemistry and Biochemistry,
   University of California, Santa Barbara, California 93106, USA}

\newcommand{\eq}{Eq.~}
\newcommand{\fig}{Fig.~}

\newcommand{\vb}[1]{ {\bf #1}}
\newcommand{\um}{~\ensuremath{\mu}m\xspace}

\newcommand{\rb}{\ensuremath{\vb{r}}\xspace}

\newcommand{\qv}{\vb{q}\xspace}
\newcommand{\ssize}{\ensuremath{\mathcal{L}}\xspace}
\newcommand{\lsd}{\ensuremath{L_{sd}}\xspace}
\newcommand{\phiq}[1]{\ensuremath{\phi_{\bf #1}}\xspace}

\begin{document}

\setlength{\pdfpageheight}{\paperheight}
\setlength{\pdfpagewidth}{\paperwidth}


\title{Dynamic simulations of multicomponent lipid membranes over long length and time scales}
\author{Brian~A.~Camley}
\affiliation{\ucsbphysics}
\author{Frank~L.~H.~Brown}
\affiliation{\ucsbchem}
\affiliation{\ucsbphysics}

\begin{abstract}
We present a stochastic phase-field model for multicomponent lipid bilayers that explicitly accounts
for the quasi-two-dimensional hydrodynamic environment unique to a thin fluid membrane immersed in aqueous solution.  Dynamics over 
a wide range of length scales  (from nanometers to microns) for durations up to seconds and longer are readily accessed and
provide a direct comparison to fluorescence microscopy measurements in ternary lipid/cholesterol mixtures.  Simulations of phase separation
kinetics agree with experiment and elucidate the importance of hydrodynamics in the coarsening process. 
\end{abstract}


\maketitle

Multicomponent lipid bilayer membranes are of universal biological importance and are increasingly viewed as fundamentally
interesting soft matter systems \cite{edidin_raft_review,simonsvazrafts2004,greenbook}.  Ternary mixtures of saturated and unsaturated lipids and cholesterol have become a standard experimental model to study the dynamics of inhomogeneous membranes under controlled laboratory conditions \cite{keller_review,Berkowitz_BBA}.  These dynamics include: diffusion of lipid domains \cite{cicuta2007}, ``flickering" fluctuations of domain boundaries \cite{espositobaumgart2007} and phase separation kinetics \cite{veatchkeller2003}.  Analytical calculations for domain diffusion and flickering in certain regimes have helped to illuminate the relevant physics \cite{stonemcconnell95,hpw81}, which crucially depends on the ``quasi-two-dimensional" (quasi-2D) \cite{oppenheimerdiamant2009} hydrodynamic environment of a viscous fluid membrane suspended within
bulk solvent \cite{saffmandelbruck75,lubenskygoldstein96}.

Particle based simulations have been used to study general features related to phase separation dynamics in bilayer systems
\cite{mcwhirter2004,kumar05,gompper10}, but it remains difficult to directly compare simulation to experiments that
commonly involve vesicles tens of microns in diameter and observation times that run up to minutes.  At a coarser level, phase-field
representations of inhomogeneous membranes have been introduced \cite{thornton07,lindenberg08,Fan2010}, but these studies have neglected hydrodynamic
effects and/or thermal fluctuations, again making detailed comparison to experimental dynamics impossible.   This letter
presents a simulation scheme that combines quasi-2D hydrodynamics and thermal fluctuations with the capability to probe experimentally
relevant length and time scales.  Compelling agreement with both theory and experiment is obtained, suggesting this methodology as a powerful tool
to study membrane dynamics in the context of physics, biophysics and biology.

Our focus is on experimentally observable dynamics, not first-principles prediction 
of detailed thermodynamic properties.  Accordingly, we adopt a standard Landau-Ginzburg free energy functional
for binary mixtures \cite{safran,chaikinlubensky}
\begin{equation}
F = \int d^2 \vb{r} \left[ -\frac{r}{2} \phi^2 + \frac{u}{4} \phi^4 + \frac{\gamma}{2} \left| \nabla \phi \right|^2 \right]
\label{eq:lg_F}
\end{equation}
where $r,u,\gamma>0$.  \eq \ref{eq:lg_F} may be interpreted as a phenomenological description of the observed two-phase coexistence in ternary lipid/cholesterol systems.  However, to facilitate 
comparison with experiment, we assume tight stoichiometric complexation between cholesterol and the saturated lipid\cite{mcconnell2005} and
identify $\phi(\vb{r}) \equiv \chi(\vb{r})_{\textrm{unsaturated}}- \chi(\vb{r})_{\textrm{complex}}$ as the local
difference in mole fraction ($\chi$) between lipid species.  
Though limited to mixtures with a 1:1 stoichometry between saturated lipids
and cholesterol, this picture has the advantage of simplicity: only a single composition field is required and the three parameters appearing 
in $F$ are readily related to measurable experimental properties \cite{safran}: the line tension between phases, $\sigma = \frac{2\sqrt{2}}{3 u} \gamma^{1/2} r^{3/2}$, interface width $\xi = \frac{1}{2}\sqrt{\gamma/r}$, and equilibrium phase compositions $\phi_0 = \pm \sqrt{r/u}$.

We require the dynamics of $\phi$ to conserve lipid concentrations, hydrodynamically couple 
points on the membrane and have thermal fluctuations.  Model H dynamics represents the generic long-wavelength low-frequency picture to incorporate these requirements \cite{chaikinlubensky}.
In the overdamped ``creeping-flow" limit for experimental conditions (low Reynolds number, $Re < 10^{-3}$), model H reduces to \cite{kogakawasaki91,bray94}
\begin{eqnarray}
\lefteqn{(\partial_t + \vb{v}\cdot\nabla)\phi(\rb,t) = M \nabla^2 \frac{\delta F}{\delta \phi(\rb,t)} + \theta(\rb,t)} \label{eq:modelH_phi}  \label{eq:modelH_v} \\ \nonumber
 v_i(\rb,t) &=& \int d^2r' \, T_{ij}(\rb-\rb') \left[ \frac{\delta F}{\delta \phi} \nabla_j' \phi(\rb',t) + \zeta_j(\rb',t) \right] .
\end{eqnarray}
Though typically applied to pure three dimensional (3D) or two dimensional (2D) geometries,  we
use \eq \ref{eq:modelH_v} for the quasi-2D fluid membrane geometry first introduced by Saffman
and Delbr\"uck  \cite{saffmandelbruck75}, which considers the membrane to be a thin, flat fluid surface with surface viscosity $\eta_m$
surrounded by a bulk fluid with viscosity $\eta_f$.  In this case $\vb{v}=(v_x,v_y)$ is the in-plane membrane velocity field, $M$ is a transport coefficient related to the collective diffusion coefficient for lipids within the bilayer ($D_\phi$) via $M=D_{\phi}/2r$ \cite{chaikinlubensky,mcwhirter2004}
and $T_{ij}(r)$ is the Green's function  for in-plane velocity response of the membrane to a point force \cite{lubenskygoldstein96,oppenheimerdiamant2009}.
In a conventional 3D fluid geometry, $T_{ij}(r)$ would be the Oseen tensor \cite{doiedwards}; its form for the quasi-2D membrane geometry includes the effect
of flow both within the membrane and in the bulk solvent.  There is no simple closed form expression for $T_{ij}(r)$, but its Fourier 
transform is \cite{lubenskygoldstein96,oppenheimerdiamant2009}
\begin{equation}
T_{ij}(\qv) = \frac{1}{\eta_m (q^2 + q/\lsd)} \left( \delta_{ij} - \frac{q_i q_j}{q^2} \right) \label{eq:membrane_oseen}
\end{equation}
where $\lsd = \frac{\eta_m}{2 \eta_f}$ is the Saffman-Delbr\"uck length scale.
The Gaussian white thermal noise terms, $\theta(\rb,t)$ and $\zeta_j(\rb',t)$, are distributed with variances set by the
fluctuation-dissipation theorem \cite{vankampen}.  In \eq \ref{eq:modelH_v} and henceforth the Einstein summation
convention is assumed.

In typical model membrane systems, $\eta_m$ ranges from $(0.1-10) \times 10^{-6}$ surface poise (poise-cm, or grams/s) \cite{dimova1999,petrovschwille2008,cicuta2007}, corresponding to Saffman-Delbr\"uck lengths $\lsd \sim 0.1-10$ microns.  
Eq. \ref{eq:membrane_oseen} exhibits the characteristic scaling ($1/r$ in real space) associated with the usual 3D Oseen tensor for $q\ll \lsd^{-1}$, and reduces exactly to the 2D analog to the Oseen tensor for $q\gg \lsd^{-1}$ \cite{oppenheimerdiamant2009}.  However, fluorescence microscopy experiments probe wavelengths comparable to $\lsd$ and it is critical to use the full expression in comparison to these experiments.

Koga and Kawasaki \cite{kogakawasaki91} suggested the use of fast Fourier transforms as an efficient way to numerically
evolve zero-temperature overdamped model H dynamics.  We have extended their approach to include stochastic thermal forces and the quasi-2D
hydrodynamics discussed above, evolving \eq \ref{eq:modelH_v} in Fourier space.  Our evolution uses a Stratonovich scheme \cite{kloedenplaten} with semi-implicit terms similar to those used for the deterministic Cahn-Hilliard equation \cite{zhuchenshen1999}.  This simulation methodology may be viewed as an extension of traditional ``Brownian dynamics with
hydrodynamic interactions"\cite{mccammon} to composition dynamics within a flat quasi-2D membrane environment.  Details of the numerical scheme are in the appendix.  

\begin{figure}[h]
\includegraphics[width=70mm]{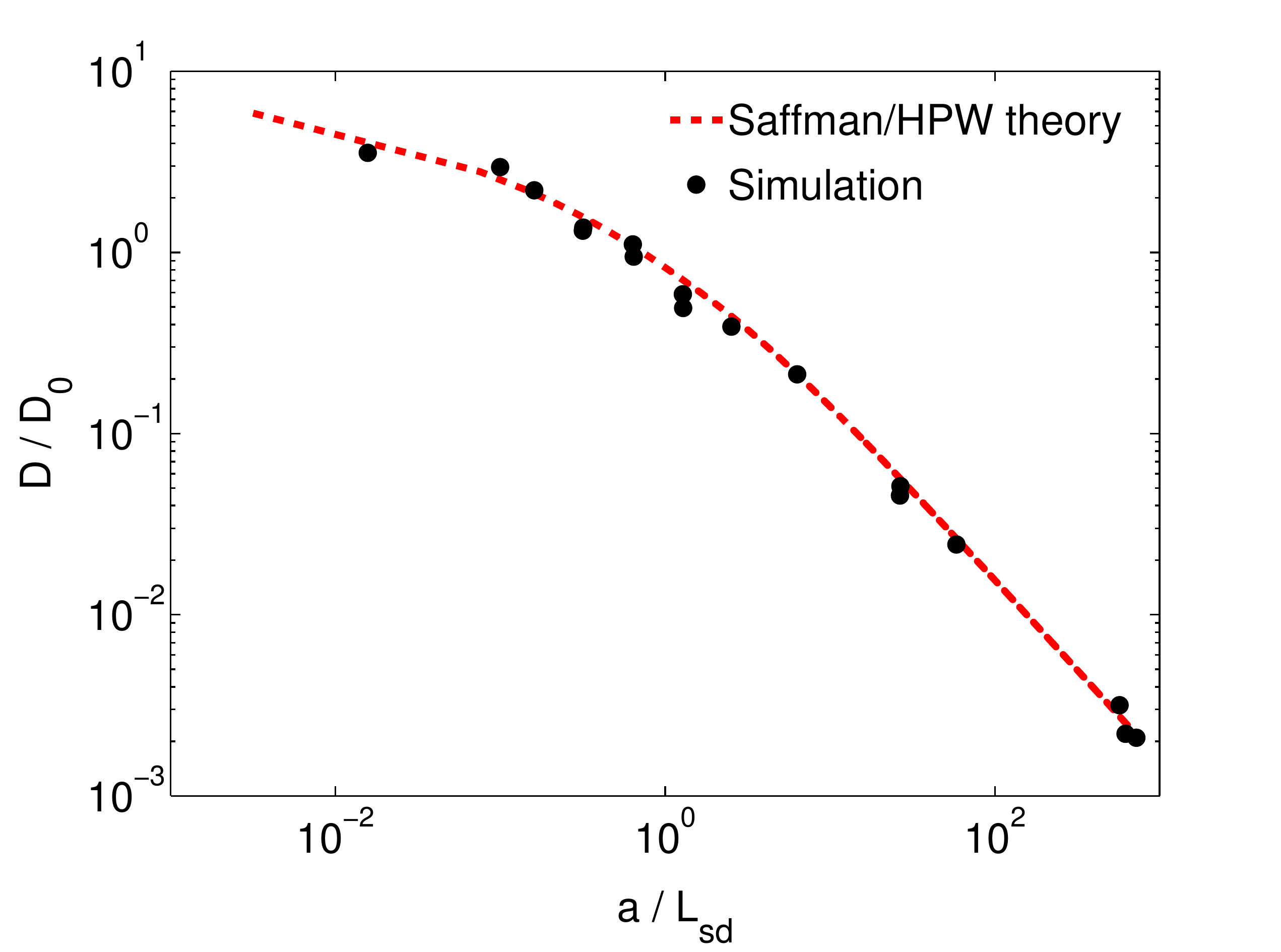}
\caption{(Color online).  Translational diffusion coefficients obtained by tracking simulated domain motion, compared with Saffman-Delbr\"uck-Hughes-Pailthorpe-White theory (using the interpolation of \cite{petrovschwille2008}).  $D_0 = k_B T / 4\pi\eta_m$.  
Error bars are on the order of marker size.  Line tensions and temperatures were chosen to ensure domains remained circular over the course of the simulations; box sizes range from 10 $\mu$m to 40 $\mu$m (N = 256 to 1024), chosen to eliminate effects from periodic boundary conditions.}
\label{fig:diffusion}
\end{figure}

The translational diffusion of circular lipid domains presents an ideal test to assess the validity of the simulation method.  Theoretical results based upon the quasi-2D
hydrodynamics underlying our approach have been derived \cite{saffmandelbruck75,hpw81} and confirmed experimentally \cite{cicuta2007,petrovschwille2008,sakuma_saffman}.
These results predict $D_a = (k_B T / 4 \pi \eta_m) \mathcal{F}(a/\lsd)$ for the diffusion coefficient of a domain of radius $a$.  The function $\mathcal{F}$ represents the solution
to an integral equation \cite{hpw81}, but is well approximated by a closed-form empirical fit described in \cite{petrovschwille2008}.
Choosing initial conditions to reflect a single circular domain and thermodynamic parameters that guarantee the domain remains nearly circular (high $\sigma$, low $T$), we track
the domain's position over time and infer $D_a$ via its mean square displacement.  The results collapse onto $\mathcal{F}(a/\lsd)$ (all of $a$, $\eta_m$ and $\eta_f$ were independently
varied) over a wide range of $a/\lsd$ ratios, completely spanning the crossover between Saffman-Delbr\"uck diffusion $D_a \sim \ln(\lsd/a)$ in the limit of small domains and Stokes-Einstein-like
diffusion $D_a \sim 1/a$ in the limit of large domains  (\fig \ref{fig:diffusion}).

\begin{figure}[h] 
\includegraphics[width=60mm]{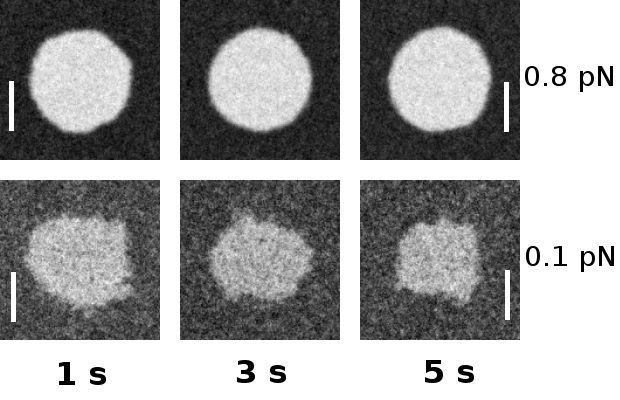}
\includegraphics[width=70mm]{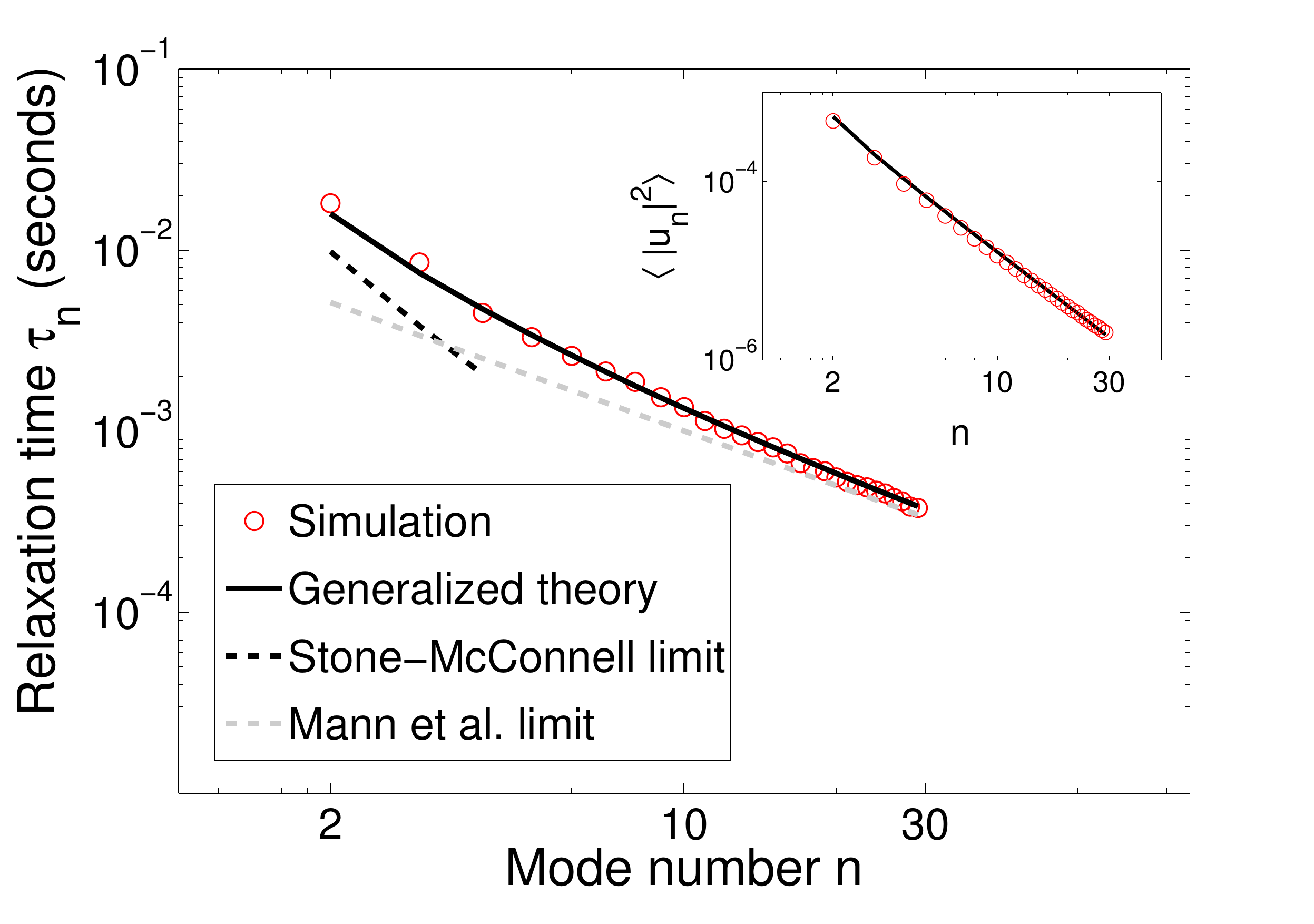}
\caption{(Color online).  TOP: Simulated evolution of a flickering domain with initial radius $R = 2.0 \mu$m, $\eta_m = 1 \times 10^{-6}$ surface poise, $\sigma = 0.8, 0.1$ pN and $\xi = 40$~nm, $\phi_0 = 0.4$.  The collective diffusion coefficient $D_\phi = \frac{k_B T}{4\pi\eta_m} = 3.2 \times 10^{-9}$ cm$^2$/s.  $N = 512$, and the time step $\Delta t$ is 5 $\mu$s. $T = 21^\circ$C, $\eta_f = 0.01$ Poise.  These are typical values for ternary domains near 20$^\circ$C, with a viscosity in the middle of the commonly accepted range (see text).  The $\sigma = 0.1$ pN domain shrinks, whereas
the $\sigma=0.8$ pN domain is stable.  The box size for the simulation is 20 $\mu$m $\times$ 20 $\mu$m (Scale bar: 2 $\mu$m.  Only the region immediately surrounding the domain is displayed.).  BOTTOM: Relaxation times for mode $n$ for $\sigma = 0.8$ pN (see text).  The theoretical predictions are summarized in \eq \ref{eq:modes}.  Inset: thermal variance in amplitude for mode $n$.  The theoretical prediction (black) is the equipartition result described in the text.  There is no fitting involved for either the main figure or inset.  Theoretical predictions follow immediately from the parameters input to the model and demonstrate the robustness of the simulation scheme.}
\label{fig:flicker}
\end{figure}

\begin{figure}[h]
\includegraphics[width=90mm]{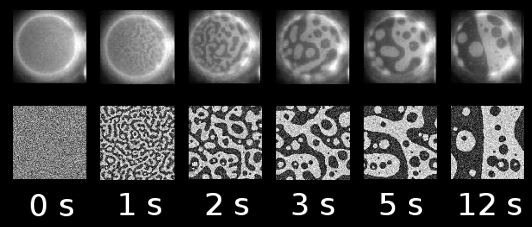}
\caption{Comparison of experimental (top) and simulated (bottom) phase separation kinetics for 1:1:1 mixtures of DOPC/DPPC/Chol.  Experimental figure adapted from Veatch and Keller (vesicle diameter is $30\um$) \cite{veatchkeller2003}.  See text for physical parameters and $\mathcal{L}=30\um$,  $N=1024$ and $\Delta t = 20 \mu$s.}
\label{fig:compare}
\end{figure}

In experimental membrane systems, line tensions are seldom high enough to fully suppress shape fluctuations of lipid domains.  Though these fluctuations have minimal 
impact on translational diffusion, the fluctuations themselves may be analyzed \cite{espositobaumgart2007} and provide a further test of our simulation methods.  Evolving an initially circular domain by Eqs. \ref{eq:modelH_phi}, we found, in qualitative agreement with \cite{frolov2006}, that domains with small line tensions $\sim0.1-0.2$~pN are not
thermodynamically stable at temperatures  $\sim 20^{\circ} C$ and area fractions $\sim 0.03-0.1$; the domain radii shrink over time in favor of a more homogeneous distribution through the simulation box (\fig \ref{fig:flicker}).  (However, this homogeneous phase does not appear to be composed of ``an ensemble of small domains" \cite{frolov2006};  the composition is rapidly fluctuating everywhere, without any clearly defined domain boundaries.)  This instability depends not only on line tension, but also on the lipid composition of the membrane; for 1:1:1 mixtures (area fraction 50\%) with similar physical properties, micron-scale phase separation is observed both numerically and experimentally (\fig \ref{fig:compare}).

At higher line tensions, domains under similar thermodynamic conditions are stable and we have simulated the dynamic fluctuations of these systems (\fig \ref{fig:flicker}) 
We use the image analysis techniques of \cite{espositobaumgart2007}, tracing the boundary of the domain, and expanding it into quasi-circular modes, $r(\theta,t) = R_o(1 + u_o(t) + \frac{1}{2}\sum_{n\neq0} u_n(t) e^{i n \theta})$.  For small deviations $u_n$, the domain shape is expected to behave in accord with an effective Hamiltonian $H = \sigma L \approx \frac{\sigma \pi R_o}{2} \sum_{n>0} (n^2-1) |u_n|^2$ ($L$ is the domain perimeter); thus
$\langle |u_n|^2\rangle = 2 k_B T / \sigma \pi R_0 (n^2-1)$ is expected via equipartition.  Experimentally, this relation is used to determine line tensions \cite{espositobaumgart2007}.  We observe that our numerical experiments yield the equipartition result with the expected $\sigma = \frac{2\sqrt{2}}{3 u} \gamma^{1/2} r^{3/2}$ (\fig \ref{fig:flicker} inset).
The linear dynamics of these modes depend on the line tension as well as the viscosities of both the membrane and the surrounding fluid; $\langle u_n(t) u_{-n}(0) \rangle = \langle|u_n|^2\rangle e^{-t/\tau_n}$.
Stone and McConnell found these relaxation times neglecting the membrane viscosity (valid for length scales $R/n \gg \lsd$) and Mann et al. calculated them neglecting the bulk fluid (valid for $R/n \ll \lsd$) \cite{stonemcconnell95,mann1995}.  In our simulations, we see exponential relaxation over all modes $u_n(t)$ with $\tau_n$ plotted in \fig \ref{fig:flicker}.  We note that for large $n$ modes, the Mann theory is appropriate, as expected, but there are deviations from the Stone-McConnell result even at low $n$ since $\lsd= 0.5 \mu$m is
comparable to the size of the domain.  The simulations are in excellent agreement with a recent generalization to the Stone-McConnell theory \cite{camleyesposito2010} 
that includes the effects of both $\eta_m$ and $\eta_f$.  For reference, we include the three expressions here:
\begin{eqnarray}
\label{eq:modes}
\tau_n^{\textrm{Stone}} &=& \frac{2\pi R_o^2 \eta_f}{\sigma} \frac{n^2-1/4}{n^2(n^2-1)} ; \tau_n^{\textrm{Mann}} =  \frac{4 \eta_m R_o}{n \sigma}  \\
\tau_n^{\textrm{general}} &=& \frac{\eta_m R_o}{\sigma} \frac{1}{n^2(n^2-1)} \left[\int_{0}^{\infty}dx \frac{J_n^2(x)}{x^2(x+R_o/\lsd)}\right]^{-1}. \nonumber
\end{eqnarray}
where $J_n(x)$ is a Bessel function of the 1st kind.

The modeling described herein displays its true potential when applied to problems that are not easily explained with analytical theory.  Using identical methodology to the
diffusion and fluctuation studies, but employing a homogeneous initial condition, allows the study of phase separation kinetics in ternary lipid/cholesterol 
model systems.  Our simulations are motivated by the experiments of Veatch and Keller \cite{veatchkeller2003} on roughly 1:1:1 mixtures of DOPC/DPPC/Chol.  
Taking care to choose parameters consistent with the experimental system, we find strong qualitative similarity between simulation and experiment (\fig \ref{fig:compare}).  
However, we stress that this comparison has limitations.  The experimental temperature quench taken in \cite{veatchkeller2003} was not
precisely controlled or recorded; our homogeneous starting point, followed by constant $T$ dynamics represents a numerically convenient choice, adopted 
in the absence of clear experimental guidance.   A further  limitation is that we have assumed both phases share the same viscosity.  Lipid diffusion coefficients in ordered and disordered phases differ roughly by a factor of ten \cite{korlach1999}, suggesting a comparable difference in viscosities between the two phases.  The adoption of a single viscosity 
for the two phases is an approximation; nevertheless, certain composition dynamics in ternary model systems 
appear to be adequately described using single-viscosity theories employing (either explicitly or implicitly) a single ``effective viscosity'' for the membrane \cite{camleyesposito2010,petrovschwille2008,cicuta2007}.  This approximation may not be as severe as it first appears.

The model uses eight physical parameters  ($T$,$\sigma$,$\xi$,$\phi_0$,$D_{\phi}$,$\eta_f$,$\eta_m$,$\mathcal{L}$).  The temperature $T=21^{\circ}C$, 
box size $\mathcal{L}=30\um$ and viscosity of water $\eta_f = 0.01$ Poise follow immediately from the experimental conditions.  The remaining five parameters
specify details of the specific bilayer system under study.  Though precise measurements of all these parameters are not available, the numbers 
are known approximately, either for the specific DOPC/DPPC/Chol system under study or by analogy to other model systems expected to show similar behavior.
$\phi_0=0.4$ may be determined from the DOPC/DPPC/Chol phase diagram \cite{cicuta2007}.  
The line tension $\sigma=0.1$~pN is based on flicker spectroscopy measurements of DOPC/DPPC/Chol mixtures \cite{espositobaumgart2007}, 
which also sets the rough order of the correlation length $\xi=40$ nm via the relation $\sigma\xi \approx k_B T$, a result
motivated by the behavior of the Ising model  \cite{exactlysolved} and believed to be consistent with experiments in ternary lipid systems \cite{honerkampsmith2008}.  
The transport coefficients
$\eta_m=5\times 10^{-6}$ surface Poise and $D_{\phi}=7 \times 10^{-10}$ cm$^2$/s are difficult to precisely measure and are not known experimentally
for the exact system studied here.  The values chosen are consistent with measurements in related lipid systems \cite{korlach1999,petrovschwille2008}
and the requirement that $D_{\phi} \approx \frac{k_B T}{4 \pi \eta_m}$ \cite{saffmandelbruck75}.

The similarity between experiment and theory in \fig \ref{fig:compare} is striking and demonstrates that complex nonlinear bilayer dynamics over micron length scales
and second time scales may be directly studied numerically.  
The displayed results depend upon both membrane and solvent viscosity; a naive 2D simulation assumes the wrong dissipation for length scales above \lsd ($\sim 1 \mu$m).   
More interestingly, dynamics of quasi-2D 
membrane phase separation are qualitatively different from the pure 2D case.  We observe dynamical scaling with a continous morphology, with length scale $R(t) \sim t^{1/2}$ for critical mixtures (1:1:1) at $R(t) \gg \lsd$, which
can be explained using simple scaling arguments (as in \cite{bray94}).  In pure 2D hydrodynamic systems in the creeping flow limit, dynamical scale 
invariance breaks down \cite{wagneryeomans98}; in the absence of hydrodynamics, $R(t)\sim t^{1/3}$ \cite{bray94,wagneryeomans98,zhuchenshen1999}.  $R(t) \sim t^{1/2}$ scaling was also seen in \cite{kumar05}; our results show that this scaling emerges naturally from quasi-2D hydrodynamics. Details of the unique scaling
aspects of quasi-2D phase separation kinetics will be presented in a forthcoming paper.

Our modeling is consistent with experimental \cite{cicuta2007,sakuma_saffman,petrovschwille2008,camleyesposito2010,veatchkeller2003} and theoretical results \cite{saffmandelbruck75,mann1995,stonemcconnell95,camleyesposito2010} for a wide variety of phenomena in membrane biophysics.    This approach lays the foundation for many extensions to more complex membrane and biomembrane systems.  Our approach is readily combined with continuum simulations of out-of-plane membrane undulations \cite{linbrown2004} and coupling to the cytoskeleton \cite{linbrown2005}.  The fluctuating-hydrodynamics scheme we have used is perfectly suited for an ``immersed-boundary" \cite{atzberger2007} treatment of integral membrane proteins \cite{naji}, in which stochastic fluid velocities (e.g. \eq \ref{eq:modelH_phi}) are directly coupled to protein motion.  This sort of simulation may act as a  bridge, using dynamics verified in model membrane systems to elucidate our understanding of biomembranes.  
In particular, it will be interesting to further investigate both the thermodynamic nature
\cite{frolov2006} and dynamics \cite{Fan2010} of lipid raft models in light of the results obtained in this work.

This work was supported in part by the NSF (grant nos. CHE-0848809, CHE-0321368) and the BSF (grant no. 2006285).  B.A.C. acknowledges the support of the Fannie and John Hertz Foundation.  We thank T. Baumgart, H. Diamant and S. Keller for helpful discussions.

\appendix

\section{Numerical details of simulation method}

Our overdamped model H for simulations of phase separation in a model membrane is given by
\begin{eqnarray}
\lefteqn{(\partial_t + \vb{v}\cdot\nabla)\phi(\rb,t) = M \nabla^2 \frac{\delta F}{\delta \phi(\rb,t)} + \theta(\rb,t)} \label{eq:modelH_phi_app}  \\\label{eq:modelH_v_app} 
 v_i(\rb,t) &=& \int d^2r' \, T_{ij}(\rb-\rb') \left[ \frac{\delta F}{\delta \phi(\rb',t)} \nabla_j' \phi(\rb',t) + \zeta_j(\rb',t) \right] .
\end{eqnarray}
where the continuum Fourier transform of $T_{ij}$ is \cite{lubenskygoldstein96,oppenheimerdiamant2009}
\begin{equation}
T_{ij}(\qv) = \int d^2 r \, T_{ij}(\rb) e^{-i \qv\cdot\rb}= \frac{1}{\eta_m (q^2 + q/\lsd)} \left( \delta_{ij} - \frac{q_i q_j}{q^2} \right) \label{eq:membrane_oseen_app}
\end{equation}
where the integral is over all space.  

We follow Koga and Kawasaki \cite{kogakawasaki91} in using the fast Fourier transforms (FFT) as an efficient way to numerically
evolve these dynamics.  We modify their approach by using the Oseen tensor appropriate for the quasi-2D hydrodynamic environment as well as stochastic thermal forces.  An $\mathcal{L}\times\mathcal{L}$ periodic geometry is assumed; the dynamics
of the Fourier modes $\phi_\qv = \int_0^{\mathcal{L}} \int_0^{\mathcal{L}}d^2 r \, \phi(\vb{r}) e^{-i\qv\cdot\vb{r}}$ follow from \eq \ref{eq:modelH_v_app}:
\begin{eqnarray}
 \label{eq:modelH_phi_q}
\partial_t \phiq{q}(t) + \{\vb{v}\cdot\nabla\phi(\rb,t)\}_\qv=  -M q^2 \left\{\frac{\delta F}{\delta \phi(\rb,t)}\right\}_\qv + \theta_\qv  \\  
v_{\qv,i}(t) = T_{ij}(\qv) \left\{ \frac{\delta F}{\delta \phi(\rb,t)} \nabla_j \phi(\rb,t) + \zeta_j \right\}_\qv \\
\;\;\;\: \langle \theta_\qv(t) \theta^*_{\qv'}(t') \rangle = 2 k_BT M q^2 \ssize^2 \delta_{\vb{q},\vb{q}'}\delta(t-t') \\
\langle \zeta_{\qv,i}(t) \zeta^*_{\qv',j}(t') \rangle =  2 k_B T \ssize^2 \eta_m (q^2+q/\lsd) \delta_{ij} \delta_{\vb{q},\vb{q}'}\delta(t-t').
\end{eqnarray}
where $\{f(\rb)\}_\qv$ is the Fourier transform of $f(\rb)$ and $^*$ indicates complex conjugation.  The variance of the Langevin forces $\theta$ and ${\mathbf \zeta}$ are set by the fluctuation-dissipation theorem \cite{vankampen,chaikinlubensky}.  

These equations are solved numerically by truncating to $N\times N$ Fourier modes $\qv = (m,n) 2\pi / \ssize$ with $-N/2 < m,n \le N/2$ (corresponding to a real
space discretization size $\ell = \ssize/N$).  $\{\ldots\}_\qv$ terms are evaluated in a hybrid real-space / Fourier-space fashion, handling
real-space derivatives and convolutions in $\qv$-space, local real-space operations in $\rb$-space and moving between the two representations via the FFT.

Though we have written Eqs. \ref{eq:modelH_phi_app}-\ref{eq:modelH_v_app} as two separate equations, they only represent one dynamical equation, as the velocity field is set by the composition by \eq \ref{eq:modelH_v}.  Substituting \eq \ref{eq:modelH_v_app} into \eq \ref{eq:modelH_phi_app} yields a single Langevin equation for $\phi$, but the coefficient of the thermal noise ${\mathbf \zeta}$ depends on $\phi$ through the $\nabla \phi$ term of \eq \ref{eq:modelH_phi_app}.  This so-called ``multiplicative noise" \cite{vankampen} should be treated via the Stratonovich 
interpretation, as it approximates a thermal force with a finite correlation time (set by the neglected fluid inertia) \cite{rumelin1982,vankampen}.  We use a semi-implicit Stratonovich integrator, which requires the nonconstant coefficient of the Langevin force to be averaged over its value at $\phi$ and an auxiliary value $\tilde \phi$ \cite{kloedenplaten}; the linear (but potentially most unstable) $q^4$ term is treated implicitly, and the nonlinear parts explicitly, as in semi-implicit solvers for the Cahn-Hilliard equation \cite{zhuchenshen1999}.  Our scheme is:
\begin{align}
\label{eq:evolve_line_one}\phi_\qv(t+\Delta t) &= \phi_\qv(t) - \frac{1}{1 + M \gamma q^4 \Delta t}
\left( \Delta t \left\{ \vb{v}^{\textrm{det}}\cdot \nabla \phi(t) + \vb{v}^{\textrm{therm}} \cdot (\nabla \phi(t) + \nabla \tilde \phi(t) )/2  \right. \right. \\
\nonumber &\left. \left. -M\nabla^2 (r\phi(t)-u\phi(t)^3 ) \right\}_\qv - \Theta_\qv \right) \\
v^{\textrm{det}}_{\qv,i}(t) &= T_{ij}(\qv) \left\{ \frac{\delta F}{\delta \phi} \nabla_j \phi\right\}_\qv \\
v^{\textrm{therm}}_{\qv,i}(t) &= T_{ij}(\qv) \left( \frac{1}{\Delta t}Z_{\qv,j} \right)  \\
\tilde \phi_\qv(t) &= \phi_\qv (t) + \Theta_\qv - \Delta t \left\{ \vb{v}^{\textrm{therm}} \cdot \nabla \phi(t) \right\}_\qv \\
\langle \Theta_\qv \Theta_\qv^* \rangle & =  2 k_B T M q^2 \ssize^2 \Delta t \\
\langle Z_{\qv,i} Z_{\qv,i}^* \rangle & =  2 k_B T \ssize^2 \eta_m (q^2 + q/\lsd) \Delta t \label{eq:evolve_last_line}
\end{align}

Since $\phi(\rb)$ is a real field, we know that not all modes $\phi_\qv$ are independent: $\phi_\qv^* = \phi_{-\qv}$.  This means that (assuming $N$ is even), the modes $(m,n) = (0,0),(N/2,0),(0,N/2)$, and $(N/2,N/2)$ are guaranteed to be real.  We choose these to be four of the required $N^2$ dynamical variables.  The other independent modes are chosen to be $(m,n)$ for $-N/2 < m < N/2$ and $0 < n < N/2$, $(m,0)$ for $0 < m < N/2$, $(m,N/2)$ for $0 < m < N/2$, and $(N/2,n)$ for $0<n<N/2$, as in \cite{linbrown2004}.  The real and imaginary parts of each of these modes are both independent dynamic variables.  The remaining modes are determined by the complex conjugates of the evolved modes.  We also note that because the dynamics conserves total concentration, the mode $(0,0)$ must remain constant, and is not evolved.  

The random thermal forces $\theta(\rb,t)$ and $\zeta_j(\rb,t)$ are also required to be real, which affects their Fourier transforms, and therefore the variance of the integrals $\Theta_\qv(\Delta t)$ and $Z_{\qv,j}(\Delta t)$.  We know from the variance of $\theta_\qv(t)$ (in the main paper) that $\langle \Theta_\qv(\Delta t) \Theta^*_\qv(\Delta t) \rangle = 2 T M q^2 \ssize^2 \Delta t$.  If we write $\Theta_\qv = f + i g$, we see that for the explicitly real modes $(m,n) = (0,0)$, $(N/2,0)$, $(0,N/2)$, and $(N/2,N/2)$ where $g = 0$, $\langle |f|^2 \rangle = 2 T M q^2 \ssize^2 \Delta t$, but for complex modes, $f$ and $g$ are selected from a distribution with variance $\langle |f|^2 \rangle = \langle |g|^2 \rangle = T M q^2 \ssize^2 \Delta t$.  The variance of $Z_{\qv,i}$ is exactly analogous, but with $ \langle Z_{\qv,i} Z_{\qv,j} \rangle = 2 T \ssize^2 \eta_m (q^2 + q/\lsd) \delta_{ij} \Delta t$.  

The method of Eqs. \ref{eq:evolve_line_one}-\ref{eq:evolve_last_line}
allows stable time steps to be chosen that are nearly two orders of magnitude larger than those allowed by a simple explicit scheme, comparable to the gains seen in \cite{zhuchenshen1999}.

\bibliographystyle{apsrev} 

\end{document}